\documentclass[secnumarabic,amssymb,aps,prx,reprint,superscriptaddress]{revtex4-2}
\usepackage[english]{babel}
\usepackage{graphicx}
\usepackage{amsmath}
\usepackage{amsfonts}
\usepackage[usenames,dvipsnames]{color}
\usepackage{hyperref}
\usepackage{blindtext}
\hypersetup{colorlinks}
\hypersetup{linkcolor=black, citecolor=black, urlcolor=blue}
\usepackage[utf8]{inputenc}
\usepackage{color}

\begin{document}
\title{Ionically charged topological defects in nematic fluids}
\author{Jeffrey C. Everts}
\email{jeffrey.everts@ichf.edu.pl}

\address{Faculty of Mathematics and Physics, University of Ljubljana, Jadranska 19, 1000 Ljubljana, Slovenia}
\address{Institute of Physical Chemistry, Polish Academy of Sciences, Kasprzaka 44/52, PL-01-224 Warsaw, Poland}

\author{Miha Ravnik}
\address{Faculty of Mathematics and Physics, University of Ljubljana, Jadranska 19, 1000 Ljubljana, Slovenia}
\address{Department of Condensed Matter Physics, Jozef Stefan Institute, Jamova 39, 1000 Ljubljana, Slovenia}

\date{\today}

\begin{abstract}
Charge profiles in liquid electrolytes are of crucial importance for applications, such as supercapacitors, fuel cells, batteries, or the self-assembly of particles in colloidal or biological settings. However, creating localised (screened) charge profiles in the bulk of such electrolytes, generally requires the presence of surfaces -- for example, provided by colloidal particles or outer surfaces of the material -- which poses a fundamental constraint on the material design. Here, we show  that topological defects in nematic electrolytes can perform as regions for local charge separation, forming charged defect cores and in some geometries even electric multilayers, as opposed to the electric double layers found in isotropic electrolytes. Using a Landau-de Gennes-Poisson-Boltzmann theoretical framework, we show that ions highly effectively couple with the topological defect cores via ion solvability, and with the local director-field distortions of the defects via flexoelectricity. The defect charging is shown for different defect types --- lines, points, and walls --- using geometries of ionically screened flat isotropic-nematic interfaces, radial hedgehog point defects and half-integer wedge disclinations in the bulk and as stabilised by (charged) colloidal particles.   More generally, our findings are relevant for possible applications where topological defects act as diffuse ionic capacitors or as ionic charge carriers.
\end{abstract}

\maketitle
\section{Introduction}
The ability to spatially control electric charge has relevance in a range of fields -- from charged polymers \cite{Trizac:2016, Zhao:2019}, and biological \cite{Jacobson:2017} and active matter \cite{Sindoro:2018}, to colloidal materials \cite{Dijkstra:2006}, complex fluids \cite{Shakeel:2019}, and even microelectronics \cite{Liu:2019}. Of special importance are systems where the charge carriers are ions, the so-called electrolytes; they occur in applications such as fuel cells, batteries, actuators, and sensors, but also in the biological context in cells (aqueous electrolytes). Electrolytes can be based on liquids \cite{Sun:2019}, polymer gels \cite{Gao:2018}, or solids \cite{Zhao:2020}, and depending on the application, physical properties such as high ionic conductivity or specific electrode compatibility are desired. However, importantly, the ability to locally control the charge or charge profiles in these types of electrolytes is  limited in the sense that separating charges -- i.e., creating electric double layers -- generally requires the presence of an interface, such as a solid-liquid or a liquid-liquid interface.
Here, we show that so-called nematic (liquid-crystalline) electrolytes are free of these limitations because of their partially ordered (orientational) structure.

The coupling of ionic charges to orientational (tensorial) order in nematic electrolytes, as we shall show, has some analogy with the ionic coupling to the composition (scalar) profile in mixtures of two (or more) partially immiscible fluids, such as oil and water, where the charge accumulates at the interface between the different components~\cite{Tsori:2007, Bier:2011, Bier:2012, Maciolek:2012, Tasios:2017, Everts:2017}. 
The difference of ion solvability between the two phases in such fluid-fluid mixtures, or ion partitioning, gives rise to the  Donnan (or Galvani) potential between the oil and water bulk phases and also electrifies the oil-water interface by the formation of a back-to-back electric double layer at the oil-water interface \cite{Luo:2016, Everts:2016}, mirroring the depletion region of semiconductor PN junctions. Consequently, the fluid-fluid interfacial tension is altered \cite{Bier:2008}, and the interface carries an intrinsic capacitance, which can alter the total capacitance of electrochemical cells \cite{Silva:1994, Girault:2012}. These approaches were shown recently to even drive surface phase transitions between various types of electric double layers containing an antagonistic salt \cite{Onuki:2017}. However, there is limited control for the positioning of an oil-water interface, and therefore, for where the double layer forms. In contrast, nematic fluids, because of their increased degree of order compared to isotropic fluids while still being fluidlike, offer more possibilities for charge control.

\begin{figure}[b]
\centering
\includegraphics[width=0.45\textwidth]{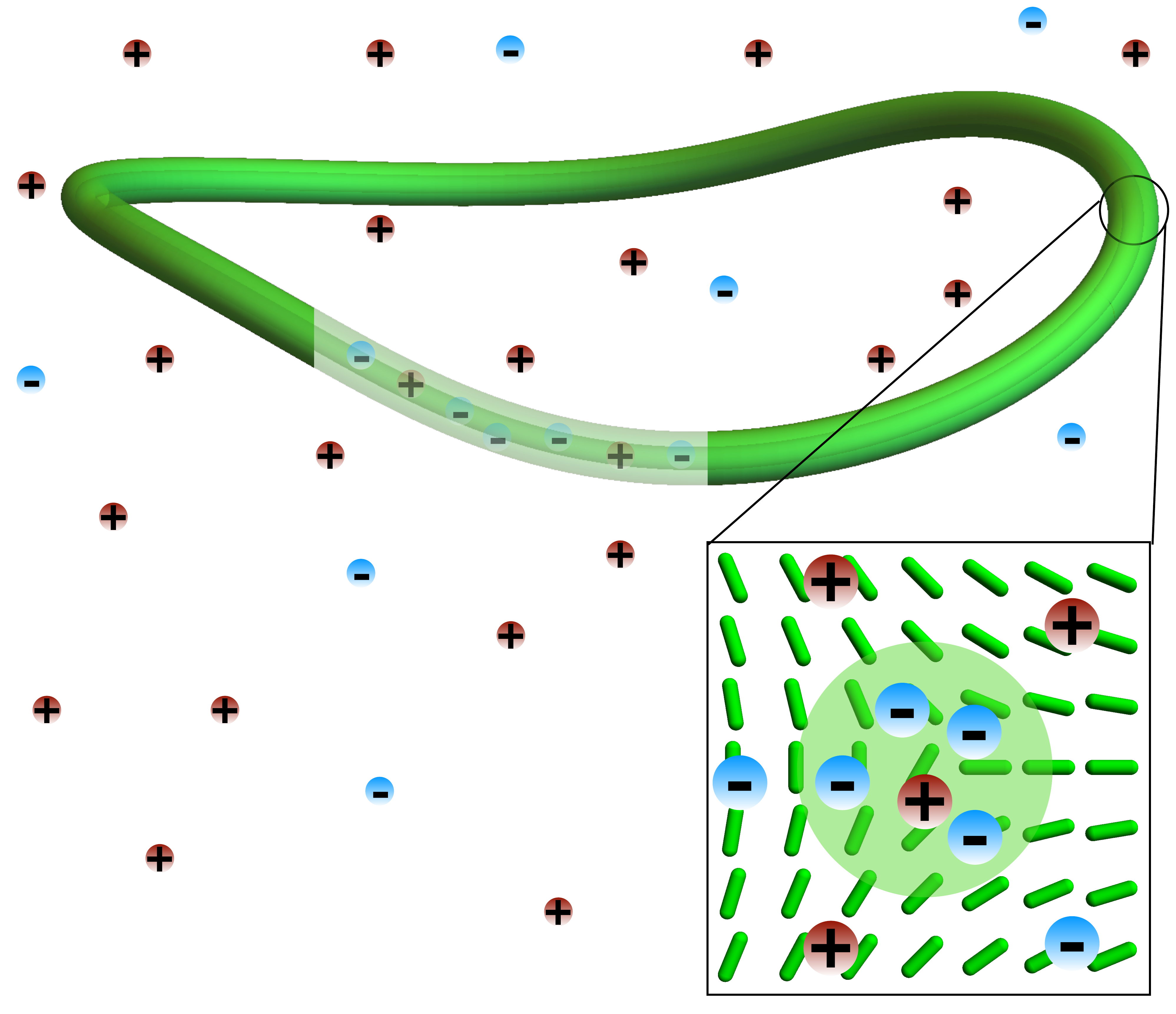}
\caption{Scheme of a nematic topological defect loop, with the isotropic core visualised in green, where positive and negative ions redistribute according to the local order parameter, director field, and flexoelectric polarisation. }
\label{fig:scheme}
\end{figure}

The manipulation of the electric charge --- typically as free ions --- was explored in nematic fluids for various motivations. A known example was the first LCD display that used ionic doping \cite{Zanoni:1968}. More recently, charge and ions have been shown to affect ion transport \cite{Neyts:2003, Neyts:2004, Barbero:2006, Poddar:2017}, surface anchoring \cite{Barbero:1993, Shah:2001, Anchor:2020}, colloidal self-assembly \cite{Mundoor:2016, Mundoor:2019, Everts:2021}, and surface charge control \cite{Ravnik:2020}. Nematic fluids possess orientational order of their building blocks, which can exhibit topologically protected patterns, called topological defects. Topological defects in the form of points, lines, and walls can be ascribed different topological invariants, including winding number and topological charge (different from electric). In recent years, the control of nematic topological defects advanced to the level that they can be designed and created into topological elements as diverse as loops, points, solitons, and even knots \cite{Smalyukh:2020, Li:2018, Pollard:2019, rahimi2017segregation, Solodkov:2019, Tran:2017, Yoon:2017}, establishing an effective, topological, soft-matter platform capable of diverse topological manipulations on the microscopic level, including with optical tweezers \cite{Musevic:2017}, external fields \cite{Noh:2016, Kim:2020, Neill:2020}, and even driving with active matter \cite{Kos:2018}. The central idea of this paper is that  efficient charge separation in the bulk of the material, combined with the possibility to manipulate the topological defects of nematic electrolytes, can then lead to the capacity to use topological defects as controllable fluidlike microelectronic elements.

In this paper, we demonstrate the use of nematic topological defects for bulk manipulation of free ions  using mesoscopic numerical modeling based on a Landau-de Gennes-Poisson-Boltzmann approach. Specifically, we show that  coupling of ions to the nematic-ordered structure is particularly strong in topological defects and leads to charging of the defect cores, which are then screened by bulk electric ``multi" (double) layers. Selected topological defect geometries are used to demonstrate the concept of electric charging of topological defects,  including point hedgehog, $\pm 1/2$ wedge defect lines, and (Saturn-ring) defect loops around spherical colloidal particles. The charging mechanism of the topological defects is shown to be governed by ion partitioning and coupling to flexoelectricity and order electricity, which can be understood from the simplified geometry of a flat isotropic-nematic interface.  More generally, the idea of this paper is that topological defects as part of a more general, topological, soft-matter platform could be transformed into soft microelectronic circuits.

\section{Methods and material system}
\label{sec:freeenergy}
We consider systems where the orientational order of a nematic electrolyte couples to ionic degrees of freedom, as schematically shown in Fig.~\ref{fig:scheme}. The exemplary materials are ion-doped nematic electrolytes, where the medium can be either thermotropic or lyotropic, with dielectric anisotropy \cite{Sonin:1987} and flexoelectricity \cite{Nguyen:2013}. A strong approach to explore topological defects at the mesoscale is by constructing the total free energy of the system $\mathcal{F}$ from different elementary contributions \cite{Gennes:1993, Roij:2010},
\begin{equation}
\mathcal{F}[\phi,\rho_\pm,{\bf Q}]=\mathcal{F}_\text{LC}[{\bf Q}]+\mathcal{F}_\mathrm{S}[\rho_\pm]+\mathcal{F}_\mathrm{C}[\rho_\pm,{\bf Q}]+\mathcal{F}_\mathrm{EL}[\phi,\rho_\pm,{\bf Q}]  ,  
\end{equation}
with $\phi({\bf r})/(\beta q_e)$ being the electrostatic potential, $q_e$ the elementary charge, and $\beta^{-1}=k_BT$ the thermal energy. We denote the ionic number densities of cations (anions) by $\rho_+({\bf r})$ ($\rho_-({\bf r})$), and the nematic-order parameter tensor with ${\bf Q}({\bf r})$. We specify each contribution below.

The free energy of a distorted --elastically or by a variable nematic degree of order-- nematic electrolyte is described by the Landau-de Gennes free energy \cite{Gennes:1993}:
\begin{align}
\mathcal{F}_\mathrm{LC}[{\bf Q}]=&\int d{\bf r}\, \Bigg\{\frac{L}{2}\partial_k Q_{ij}({\bf r})\partial_k Q_{ij}({\bf r})+ \\
&\frac{A}{2}\mathrm{tr}[{{\bf Q}({\bf r})^2}]+\frac{B}{3}\mathrm{tr}[{\bf Q}({\bf r})^3]+\frac{C}{4}\{\mathrm{tr}[{\bf Q}({\bf r})^2]\}^2\Bigg\}, \nonumber
\end{align}
with $L$ the single elastic constant, and $A$, $B$, and $C$ Landau-de Gennes bulk parameters. The single elastic constant is used for simplification, but we also note that many nematic materials have three different elastic constants of similar values. 

The nonelectrostatic part of the ions is modeled as an ideal gas contribution to the free energy \cite{Hansen:1990},
\begin{equation}
\beta\mathcal{F}_\text{S}[\rho_\pm]=\sum_{\alpha=\pm}\int d{\bf r}\, \rho_\alpha({\bf r})\{\ln[\rho_\alpha({\bf r})\Lambda_\alpha^3]-1\},
\end{equation}
{ with $\Lambda_\pm$ being the thermal de Broglie wavelength for cations and ions, respectively.} The ions are electrostatically coupled to each other and to ${\bf Q}$ via a dielectric (tensorial) coupling and via flexoelectricity and order electricity,
\begin{align}
\beta\mathcal{F}_\text{EL}[\phi,\rho_\pm,{\bf Q}]=&\int d{\bf r}\Bigg[q({\bf r})\phi({\bf r})+\nabla\phi({\bf r})\cdot{\bf P}_f({\bf Q}({\bf r})) \nonumber \\
&-\frac{1}{8\pi\lambda_B\bar{\epsilon}}\epsilon_{ij}({\bf Q}({\bf r}))\partial_i\phi({\bf r})\partial_j\phi({\bf r})\Bigg],
\end{align}
with the flexoelectric and order-electric polarization $q_e{\bf P}_f$ given in the one-constant approximation, 
\begin{equation}
({P}_f)_i({\bf r})=G\partial_jQ_{ij}({\bf r}), \quad
\label{eq:flexoQ}
\end{equation}
with $G$ the flexoelectric constant. Note that the flexoelectric and order-electric modes are mixed in the one-constant approximation; henceforth, when we mention flexoelectricity, it is implied that we mean flexoelectricity \emph{and} order electricity, unless stated otherwise. The different flexoelectric and order-electric modes can be independently tuned, if higher-order terms in gradients of $\bf Q$ are considered \cite{Ionescu:1993}. The ideal gas formulation of the nonelectrostatic part of the ionic free energy is based on selected assumptions: Ion concentrations are sufficiently low, so effects such as Bjerrum pair formation \cite{Valeriani:2010} and ion packing \cite{Orland:1997} can be neglected. For the electrostatic part, we assume that dipolar effects \cite{Orland:2007} are sufficiently weak such that only the minimal couplings are needed, and mesogens are not net charged \cite{Kondrat:2010}.
The dielectric tensor in nematic electrolytes is given by \cite{Gennes:1993} 
$\epsilon_{ij}({\bf r})=\bar{\epsilon}\delta_{ij}+\frac{2}{3}\epsilon_m^aQ_{ij}({\bf r}),$
with $\bar{\epsilon}$ the isotropic dielectric constant and $\epsilon_m^a$ the molecular dielectric anisotropy. Finally, we introduce the isotropic Bjerrum length $\lambda_B=\beta q_e^2/(4\pi\epsilon_0\bar{\epsilon})$, with $\epsilon_0$ the vacuum permittivity.

The solvation-energy {(or ion-partitioning)} contribution $\mathcal{F}_\mathrm{C}[\rho_\pm,{\bf Q}]$ is a nonelectrostatic coupling of ${\bf Q}({\bf r})$ with $\rho_\pm({\bf r})$,
\begin{equation}
\beta\mathcal{F}_\mathrm{C}[\rho_\pm,{\bf Q}]=\sum_{\alpha=\pm}\int d{\bf r}\, g_\alpha\rho_\alpha({\bf r})\mathrm{tr}[{\bf Q}({\bf r})^2].
\end{equation}
which can be interpreted as a coupling of $\rho_\pm({\bf r})$ to an external potential $\beta V_\pm({\bf r})=g_\pm\text{tr}[{\bf Q}({\bf r})^2]$, with $g_\pm$ the dimensionless Gibbs transfer energies \cite{Gros:1978, Kakiuchi:1996}. The $g_\pm$ can be interpreted as the free-energy cost for an ion to be transferred from the isotropic phase to the nematic phase, and it can be experimentally determined \cite{Jensen:2002}. A similar contribution to the free energy has been investigated in lyotropic liquid crystals \cite{Giesselmann:2005}, changing the isotropic-nematic transition temperature. Such couplings are typical in solutes, which are usually more soluble in a more disordered phase; for example, compare the solvability of a solute in gases, with liquids, and that of solids. In binary fluid-fluid mixtures, free-energy contributions of this type are introduced, either via linear coupling \cite{Debye:1965, Onuki:2006, Samin:2012}, or via couplings based on the lattice gas \cite{Bier:2012}. 

For the results presented, we use the following material parameters, which are typical for standard nematics: $L=4\times10^{-11}$ J m$^{-1}$ $A=-0.172\times10^6$ J m$^{-3}$, $B=-2.12\times10^6$ J m$^{-3}$, $C=1.74\times10^6$ J m$^{-3}$ (with equilibrium nematic degree of order $
S_b=\frac{1}{2}(\frac{-B}{3C}+\sqrt{\left(\frac{B}{3C}\right)^2-\frac{8A}{3C}})\approx 0.53$), and dielectric properties  \cite{Bogi:2001} $\bar{\epsilon}=10.3$ (so $\lambda_B=6 \ \mathrm{nm}$), $\epsilon_\parallel^m=33.3$, $\epsilon_\perp^m=12.4$, and \mbox{$\epsilon_a^m=\epsilon_\parallel^m-\epsilon_\perp^m$}. Typical values for the flexoelectric coefficient for calamitics are \mbox{$q_eG=1-10$ pC m$^{-1}$} \cite{Murthy:1993, Castles:2012}, whereas bent-core nematics can have values in the nC m$^{-1}$ range \cite{Harden:2006}. For  solvability of ions in nematics we assume that they are dissolved better in the disordered (isotropic) phase, taking $g_+=3$ and $g_-=8$; note that the exact numbers depend on the type of the solvent, the temperature, and the type of ions. Negative values of $g_\pm$ correspond to a preference for the nematic phase, which we will not consider here but can be equally implemented. Finally, typical (isotropic) Debye screening lengths in thermotropics are between $50$ and $1000$ nm \cite{Thurston:1984, Shah:2001, Musevic:2002, Everts:2021}, and with doping, even values of about $1$ nm can be reached \cite{Ionescu:2011}. { We generally focus on regimes where Bjerrum pair formation can be neglected based on studies in isotropic electrolytes; see Fig. 4 in Ref. \cite{Valeriani:2010}. However, in order to have Debye lengths comparable to the defect size to highlight the effects of ion partitioning, in some cases we also choose densities ($\sim 10^{-4}$M) where Bjerrum pair formation could be a possibility based on Ref. \cite{Valeriani:2010}, noting that, to the best of our knowledge, Bjerrum pair formation is only poorly understood in anisotropic dielectric materials.} { For all numerical calculations performed in this work, we use the finite-element software package COMSOL Multiphysics.}

\section{Electric double layer of an isotropic-nematic interface}
\label{sec:flat}
First, we demonstrate the ion-nematic couplings in the simple geometry of a flat isotropic-nematic (IN) interface, which provides central insights into the ion coupling in the cores of nematic-electrolyte topological defects. To obtain analytical insight, we assume that the nematic order is uniaxial, 
\begin{equation}
Q_{ij}({\bf r})=\frac{3}{2}S({\bf r})\left[n_i({\bf r})n_j({\bf r})-\frac{1}{3}\delta_{ij}\right],
\end{equation}
with $S({\bf r})$ the scalar uniaxial order parameter and ${\bf n}({\bf r})$ the nematic director.
We assume ${\bf n}={\bf e}_z$ to find the Euler-Lagrange (EL) equations $\delta\mathcal{F}/\delta S(z)=0$,
\begin{align}
\frac{3}{2}\beta &LS''(z)+G\phi''(z)+\frac{\epsilon_a^m}{12\pi\lambda_B\bar{\epsilon}}[\phi'(z)]^2=  \label{eq:flatorder}\\
&3S(z)\sum_{\alpha=\pm} g_\alpha\rho_\alpha(z)+\frac{\partial}{\partial S}\left\{\frac{9}{16}\beta C[S(z)]^2[S(z)-S_\mathrm{IN}]^2\right\}, \nonumber
\end{align}
with $S_\mathrm{IN}=-2B/(9C)=0.272$, and a prime indicating a spatial derivative with respect to $z$. From $\delta\mathcal{F}/\delta \phi(z)=0$, we find the Poisson equation
\begin{align}
\left[\left(1+\frac{2}{3}\frac{{\epsilon}_a^m}{\bar{\epsilon}}S(z)\right)\phi'(z)-4\pi\lambda_BGS'(z)\right]'= \nonumber \\
-4\pi\lambda_B[\rho_+(z)-\rho_-(z)] ,
\label{eq:flatpoisson}
\end{align}
where $\delta\mathcal{F}/\delta \rho_\pm(z)=\mu_\pm$ gives
\begin{equation}
\rho_\pm(z)=\rho_s\exp\left[\mp\phi(z)-\frac{3}{2}g_\pm S(z)^2\right].
\label{eq:flationdens}
\end{equation}
The ions are grand-canonically coupled to an isotropic reservoir of total bulk ion density $2\rho_s$ and chemical potential $\beta\mu_\pm=\ln(\rho_s\Lambda_\pm^3)$. Note that another choice of the assumed uniform director field would renormalize the values of $G$ and $L$. 

The set of Eqs. \eqref{eq:flatorder}--\eqref{eq:flationdens} has selected analytical solutions: 

(i) For $\rho_s=G=0$, the analytical solution is \cite{Wittmann:2014, Everts:2016}
\begin{equation}
S(z)=\frac{S_\mathrm{IN}}{2}\left[1+\tanh\left(\frac{z}{2\xi}\right)\right],
\end{equation}
with $\xi=3\sqrt{3LC}/(-B)$ the correlation length. We fix the integration constants such that the mean order parameter equals $S_\mathrm{IN}/2$ and that, for $z>0$, we have the nematic phase, while for $z<0$, we have the isotropic phase.

(ii) For $G\neq 0$ and $\rho_s=0$, the Poisson equation can be integrated to find
\begin{equation}
\phi(z)=\frac{6\pi\lambda_BG}{\epsilon_a^m}\ln\left[1+\frac{2}{3}\frac{\epsilon_a^m}{\bar{\epsilon}}S(z)\right].
\label{eq:unscreenedorder}
\end{equation}
This result shows a potential difference $\phi_O$ between the isotropic and nematic phases, where we set the isotropic phase to zero, where
\begin{equation}
\phi_\mathrm{O}=\frac{6\pi\lambda_BG}{\epsilon_a^m}\ln\left(1+\frac{2}{3}\frac{\epsilon_a^m}{\bar{\epsilon}}S_\mathrm{IN}\right).
\end{equation}
We estimate the magnitude of this potential by using values of standard nematic electrolytes (see Sec. \ref{sec:freeenergy}) to find that $\phi_\mathrm{O}=0.2-2$. At room temperature, this result is equivalent to $5-50\ \mathrm{mV}$, --substantially large potential differences to which ion densities can couple.

\begin{figure}[t]
\centering
\includegraphics[width=0.45\textwidth]{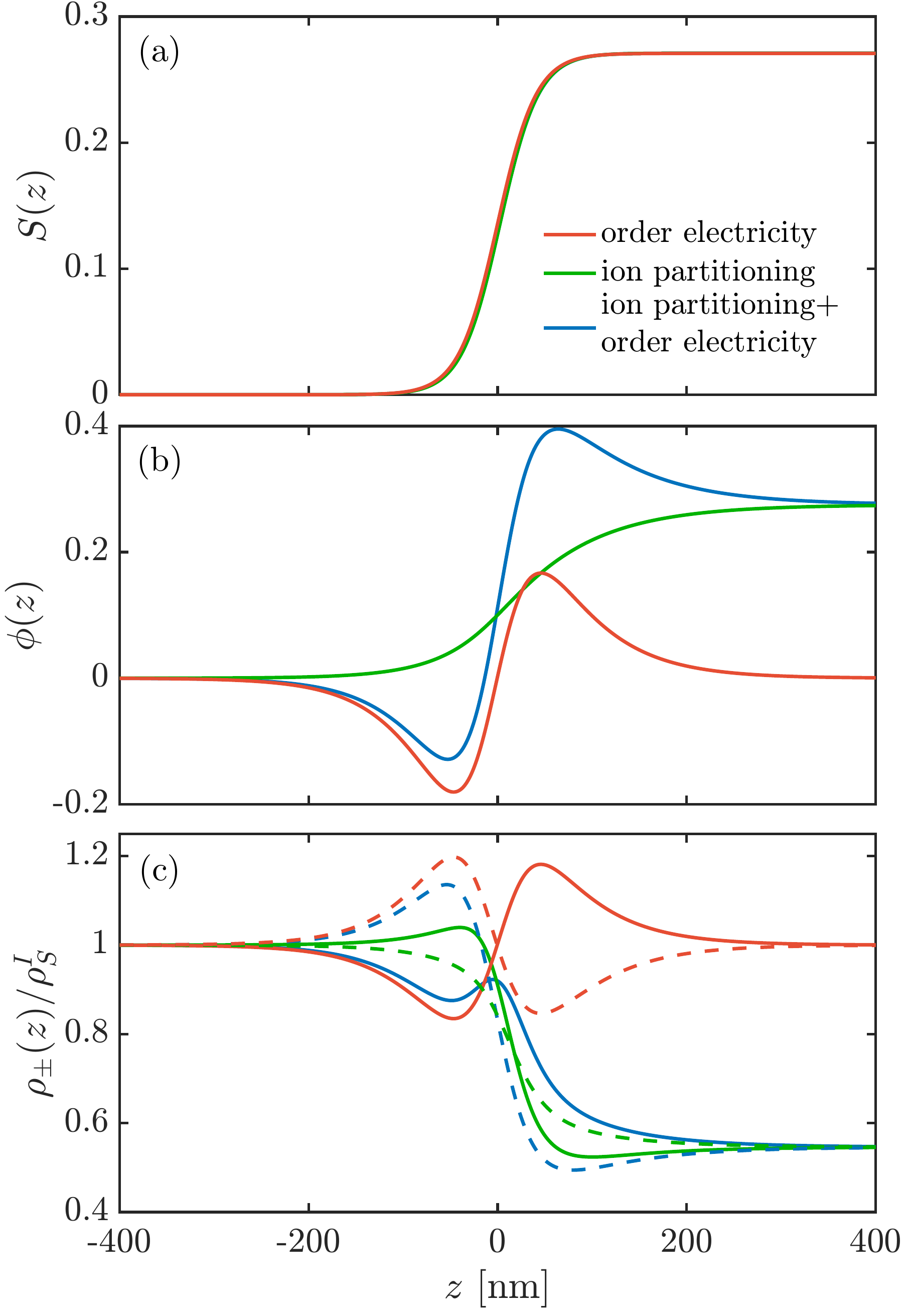}
\caption{Electric double layer of flat isotropic nematic interface (positioned at $z=0$). We show the effect of order electricity and ion partitioning on (a) the order parameter profile $S(z)$, (b) the dimensionless electrostatic potential $\phi(z)$, and (c) the ion densities for anions $\rho_-(z)$ (solid lines) and cations $\rho_+(z)$ (dashed lines). In all plots, we use $\kappa_I^{-1}=50\ \mathrm{nm}$, which results in a bulk ion density in the isotropic phase $\rho_s^I=4.4\times10^{-6}$ M, and the nematic correlation length is $\xi=20\ \mathrm{nm}$. When order electricity is included, we use the order-electric coefficient $q_eG=1\ \mathrm{pC}\ \mathrm{m}^{-1}$, and when ion partitioning is included, we use $g_+=3$ and $g_-=8$.  }
\label{fig:EDLforflat}
\end{figure}

(iii) For $G=0$ and $L\rightarrow 0$ (regime of small nematic elasticity),  the IN interface becomes sharp, i.e. $S(z)=0$ for $z<0$ and $S(z)=S_\mathrm{IN}$ for $z>0$. The Poisson equation then reduces to the modified Poisson-Boltzmann equation,
\begin{equation}
\phi''(z)=
\begin{cases}
\kappa_I^2\sinh[\phi(z)],\quad z<0, \\
\kappa_N^2\sinh[\phi(z)-\phi_\mathrm{D}], \quad z>0.
\end{cases}
\label{eq:PBBB}
\end{equation}
Here, $\phi_\mathrm{D}=(3/4)S_\mathrm{IN}^2(g_--g_+)$ is the Donnan potential. We write the Debye screening length $\kappa^{-1}_N$ ($\kappa^{-1}_I$) in the nematic (isotropic) phase, with $\kappa_i^2=8\pi\lambda_B^i\rho_s^i$, $i=I,N$ with isotropic bulk density $\rho_s^I=\rho_s$, and nematic bulk density $\rho_s^N=\rho_s\exp[-(3/4)S_\mathrm{IN}^2(g_++g_-)]$. Furthermore, we write the isotropic Bjerrum length $\lambda_B^I=\lambda_B$ and nematic Bjerrum length $\lambda_B^N=\lambda_B(\bar{\epsilon}/\epsilon_N)$, with $\epsilon_N=\bar{\epsilon}+(2/3)\epsilon_a^mS_\mathrm{IN}$. Equivalent to the boundary condition $\phi'(z)=0$ for $z\rightarrow\pm\infty$, we can impose charge neutrality of the bulk fluids, which gives us the boundary conditions $\lim_{z\rightarrow-\infty}\phi(z)=0$ and $\lim_{z\rightarrow\infty}\phi(z)=\phi_\mathrm{D}$. Together with these boundary conditions, the solution of Eq. \eqref{eq:PBBB} is equivalent to that of ions partitioning over an oil-water interface \cite{Westbroek:2015}:
\begin{equation}
\phi(z)=\begin{cases}
2\log\left[\dfrac{1+C_I\exp(\kappa_I z)}{1-C_I\exp(\kappa_I z)}\right], \quad z<0, \\
2\log\left[\dfrac{1+C_N\exp(-\kappa_N z)}{1-C_N\exp(-\kappa_N z)}\right]+\phi_\mathrm{D}, \quad z>0,
\end{cases}
\label{eq:infsharp}
\end{equation}
with integration constants $C_I$ and $C_N$. Moreover, $\phi(z)$ is continuous at $z=0$ and there is a continuity condition for the dielectric displacements, $\bar{\epsilon}\phi'(0^-)=\epsilon_N\phi'(0^+)$. If we define $\chi:=\kappa_I\bar{\epsilon}/(\kappa_N\epsilon_N)$, then the integration constants can be compactly written as
\begin{equation}
C_I=\frac{\chi+\cosh(\phi_\mathrm{D}/2)-\sqrt{1+\chi^2+2\chi\cosh(\phi_\mathrm{D}/2)}}{\sinh(\phi_\mathrm{D}/2)}
\end{equation}
and 
\begin{equation}
C_N=\frac{\sqrt{1+\chi^2+2\chi\cosh(\phi_\mathrm{D}/2)}-1-\chi\cosh(\phi_\mathrm{D}/2)}{\chi\sinh(\phi_\mathrm{D}/2)}.
\end{equation}
Finally, we find the density profiles
\begin{equation}
\rho_\pm(z)=\begin{cases}
\rho_s^I\left[\dfrac{1\mp C_I\exp(\kappa_I z)}{1\pm C_I\exp(\kappa_I z)}\right]^2, \quad z<0, \\
\rho_s^N\left[\dfrac{1\mp C_N\exp(-\kappa_N z)}{1\pm C_N\exp(-\kappa_N z)}\right]^2, \quad z>0.
\end{cases}
\end{equation}
From this analytical solution, we learn that $g_\pm$ has various effects. On the one hand, it causes a potential difference between the isotropic bulk and the nematic bulk, set by the difference $g_--g_+$. In addition, these quantities renormalize the bulk ion densities set by the sum $g_++g_-$, and, consequently, the screening length is affected. We can imagine this process as follows: A back-to-back electric double layer is formed with a higher charge density at the isotropic side of the interface, compensated by a lower, but spatially more extended charge density at the nematic side.

In addition to the analytical solutions (i) -- (iii), we numerically solve the full electric double layer at the isotropic-nematic interface, with the results presented in Fig. \ref{fig:EDLforflat}. We find that order electricity { (or dielectric anisotropy)} does not affect the interface structure, as described by $S(z)$, to a significant extent for these low salt concentrations [see Fig. \ref{fig:EDLforflat}(a), red line], { which can be understood from Eq. \eqref{eq:flatorder}, where the effective elastic torque (first term) dominates the dielectric and flexoelectric torques (second and third terms, respectively).} Furthermore, in contrast to the unscreened case [Eq. \eqref{eq:unscreenedorder}], there is no potential difference between the isotropic and nematic bulk [see Fig. \ref{fig:EDLforflat}(b), red line], although an electric field is generated close to the interface, as indicated by the modulation of the electrostatic potential around the isotropic-nematic interface. From the ion density profiles, we see that an electric double layer is formed where the bulk densities in the isotropic and nematic phases are equal [see Fig. \ref{fig:EDLforflat}(c), red lines]. Hence, the absence of a potential difference of the two bulk phases can be understood in terms of maintaining bulk charge neutrality. Finally, for $G>0$, the isotropic side of the interface is positively charged, whereas the nematic side is negatively charged, where the polarity of the double layer switches sign for $G<0$.

When only ion partitioning is included, the interface structure is not substantially affected { for these dilute systems [Fig. \ref{fig:EDLforflat}(a), green line], similar to what is known for isotropic electrolytes \cite{Onuki:2006}.} Furthermore, a Donnan potential is generated between the  isotropic and nematic phases [Fig. \ref{fig:EDLforflat}(b), green line], similar to the infinitely sharp interface [Eq. \eqref{eq:infsharp}]. Again, a back-to-back electric double layer is formed at the interface, as in the order-electric case, with the main difference being that the ion bulk densities and screening lengths in the isotropic and nematic phase are unequal [see Fig. \ref{fig:EDLforflat}(c), green line]. For this specific choice of parameters, the isotropic side of the interface is negatively charged, and the nematic side is positively charged.

When both ion partitioning and order electricity are taken into account, we see their combined effects in the electrostatic potential [Fig. \ref{fig:EDLforflat}(b), blue line]. There is a potential difference, but  a nonmonotonic modulation also appears close to the interface. In this specific case, the order-electric effect generates a double layer with opposite polarity compared to the double layer formed by ion partitioning. The result is that order electricity flips the sign of the charge compared to the  ion-partitioning-only case because we have chosen $G$ to be sufficiently large. The charge of the double layer in the order-electric case would have been enhanced when combined with ion partitioning if we had chosen $g_+>g_-$ instead of $g_->g_+$ as was done in Fig. \ref{fig:EDLforflat}.

\section{Electric double layer of a radial hedgehog topological defect}
\label{sec:hedgehog}
Topological defects in nematic electrolytes have topologically distinct director-distortion profiles, which are electrostatically susceptible to mechanisms of order electricity and flexoelectricity. An elementary topological defect in nematic electrolytes is the uniaxial radial hedgehog defect (${\bf n}={\bf e}_r$), and we demonstrate that such a topological defect can be charged with ions.
\begin{figure}[t]
\centering
\includegraphics[width=0.48\textwidth]{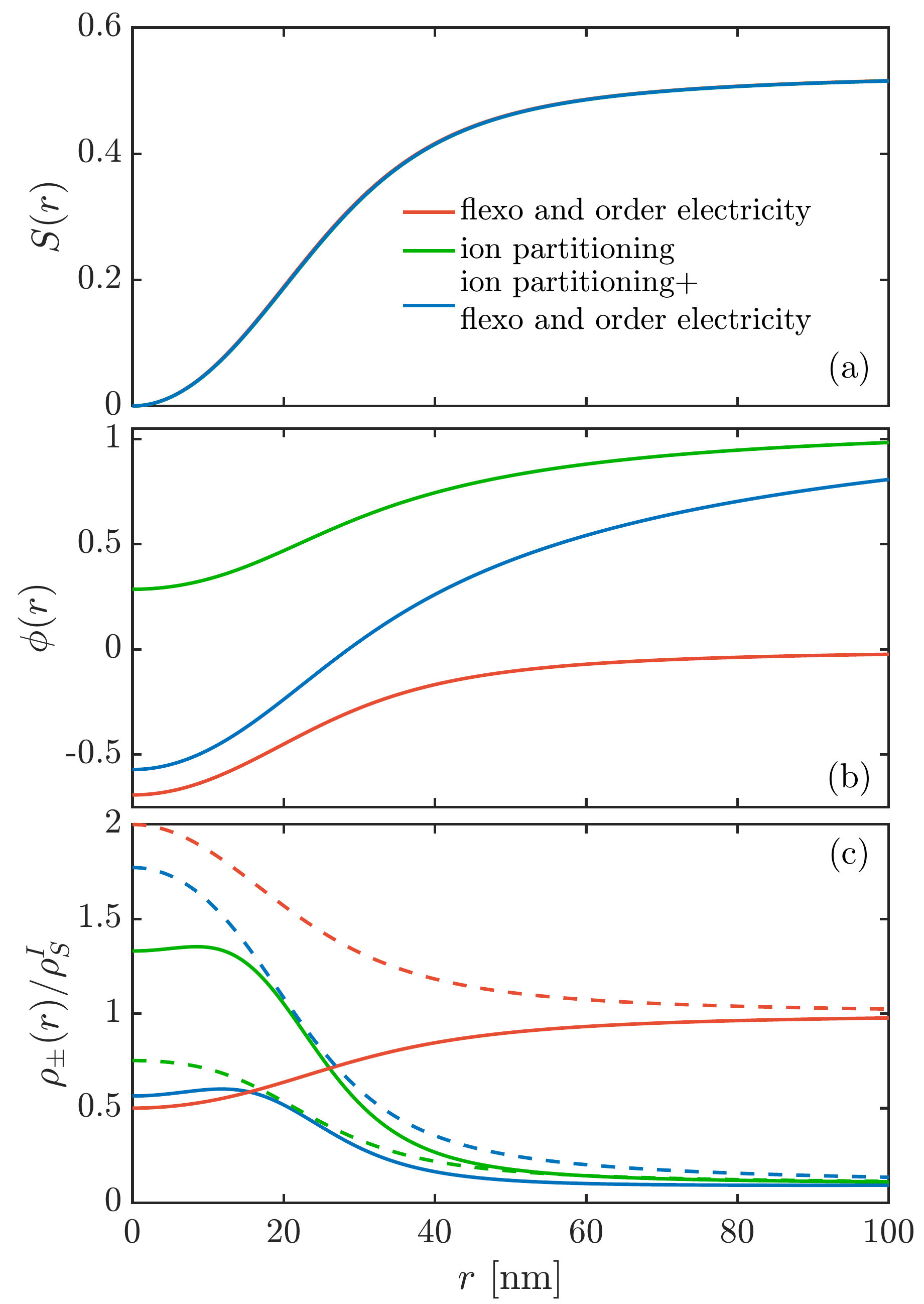}
\caption{Electric double layer of a radial $+1$ nematic topological point defect. We show the effect of flexoelectricity and order electricity, and ion partitioning on (a) the order parameter profile $S(z)$, (b) the dimensionless electrostatic potential $\phi(z)$, and (c) the ion densities for anions $\rho_-(r)$ (solid lines) and cations $\rho_+(r)$ (dashed lines). In all plots, we use $\kappa_I^{-1}=10\ \mathrm{nm}$, which results in a bulk ion density in the isotropic phase $\rho_s^I=1.1\times10^{-4}$ M. When order and flexoelectricity are included, we use $q_eG=3\ \mathrm{pC}\ \mathrm{m}^{-1}$, and when ion partitioning is included, we usee $g_+=3$ and $g_-=8$.  The system size is taken to be $R=250\ \mathrm{nm}$.}
\label{fig:EDLhedgehog}
\end{figure}
\begin{figure}[t]
\centering
\includegraphics[width=0.5\textwidth]{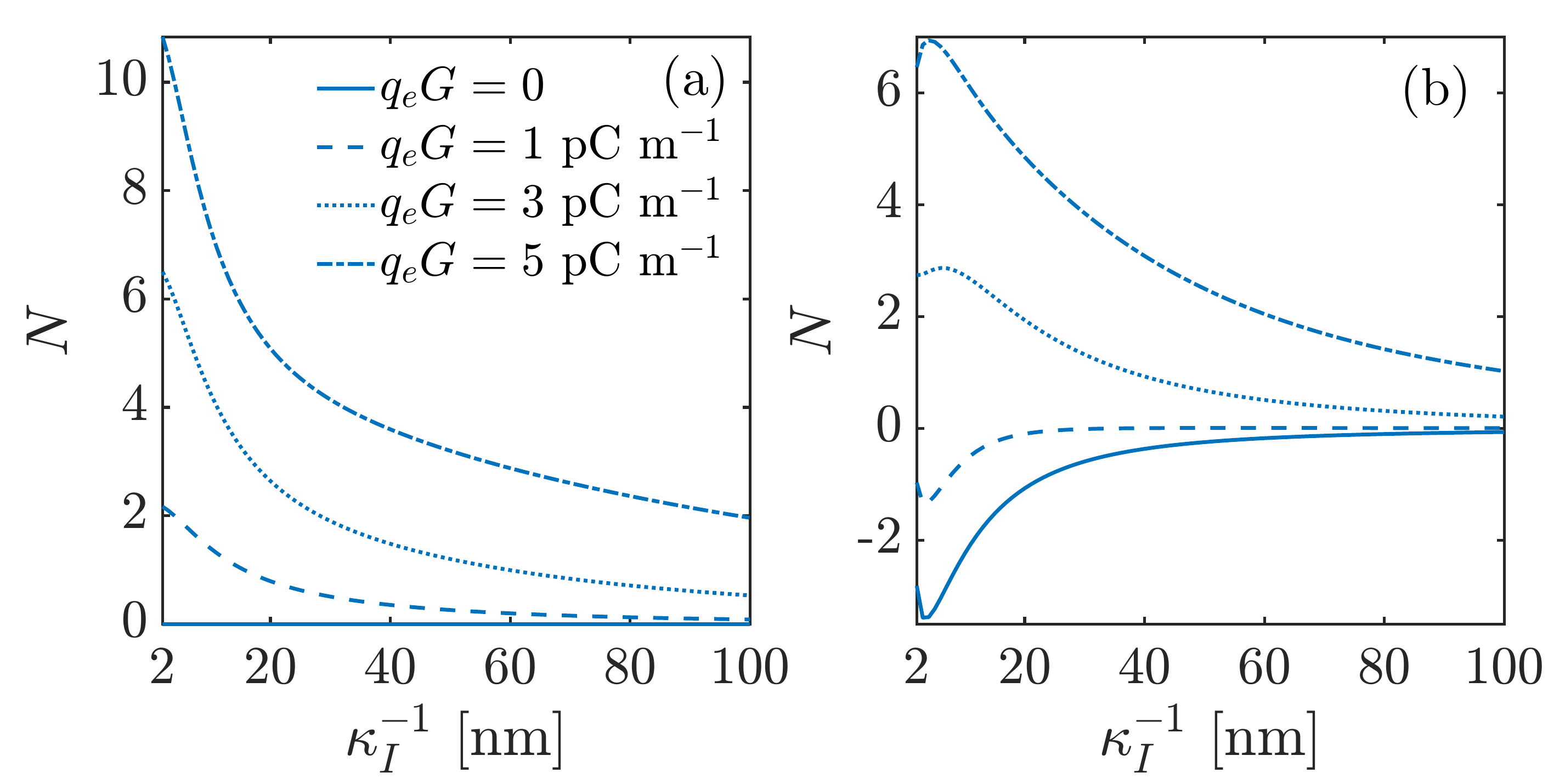}
\caption{Total charge number $N$ in the isotropic core of a radial hedgehog defect for varying values of the flexoelectric or order-electric coefficient $G$ as a function of the isotropic Debye screening length $\kappa_I^{-1}$. Panel (a) corresponds to a regime with no ion partitioning, $g_+=g_-=0$, and panel (b) corresponds to one with ion partitioning, $g_+=3$ and $g_-=8$. The system size is $250\ \mathrm{nm}$.}
\label{fig:chargenumber}
\end{figure}
The EL equations are solved using a spherical box with radius $R$, exploiting the spherical symmetry and assuming boundary conditions $S'(0)=S'(R)=0$ and  no free surface charge density. In Fig. \ref{fig:EDLhedgehog}, we show the electric-double-layer characteristics for regimes of (i) nonzero flexoelectricity $G\neq0$ and no ion partitioning $g_\pm=0$ (red curves), (ii) zero flexoelectricity $G=0$ and preferred ion solvability in the isotropic phase $g_\pm>0$ (green curves) and, (iii) nonzero flexoelectricity $G\neq 0$ and preferred ion solvability in the isotropic phase $g_\pm>0$ (blue). In line with the results for the flat interface in Sec. \ref{sec:flat}, we observe that the order parameter profile (i.e. the molten defect core) is only weakly influenced by the presence of ion partitioning or flexoelectricity and order electricity [see Fig. \ref{fig:EDLhedgehog}(a)]. As usual, we observe an isotropic core and that the order parameter attains its bulk value for $r$ sufficiently large \cite{Sluckin:1988}. 

The electrostatic potential in the defect is strongly dependent on the value of $g_\pm$ and $G$ [Fig. \ref{fig:EDLhedgehog}(b)] with the potential difference  generated between the isotropic core and the nematic bulk. The main difference from the flat isotropic-nematic interface is that the electrostatic potential at the core center does not vanish, $\phi(0)\neq 0$. This difference indicates that even when the core is perfectly isotropic, there is a net ionic charge density, as can be understood from Eq. \eqref{eq:flationdens}. In other words, the isotropic core is too small to effectively perform as an isotropic bulk, and this is also apparent in the ion density profiles, where $\rho_\pm(r)$ do not attain their bulk values $\rho_s^I$ in the isotropic core [see Fig. \ref{fig:EDLhedgehog}(c) for any of the three cases]. Actually, it is difficult to attain local charge neutrality in the isotropic phase because the interfacial width is relatively large. Oil-water droplets, in contrast, have been studied, where bulk neutrality in the droplet is achieved in Ref. \cite{Graaf:2008}, but these droplets are much larger than the isotropic core of an uniaxial hedgehog. However, we still use $\rho_s^I$ (and not $\rho_s^N$) as our reference density because then the same reference density is used for both the flexoelectric case and the ion partitioning case (recall $\rho_s^N=\rho_s^I$ for $g_\pm=0$), but they are unequal for $g_\pm\neq0$.

From Fig. \ref{fig:EDLhedgehog}(c), we see that the defect core carries a net charge. We calculate this net charge $N$ as
\begin{equation}
N=\int_{r<R_c}d{\bf r}\, [\rho_+({\bf r})-\rho_-({\bf r})],
\end{equation}
where $R_c$ is an arbitrarily chosen cutoff radius for which the system attains an order parameter of half its bulk value ($R_c\approx25\ \mathrm{nm}$). In Fig. \ref{fig:chargenumber}, we show the dependence of $N$ with respect to $\kappa_I^{-1}$ and $G$ (a) without and (b) with ion partitioning, where the partition coefficients are chosen to give rise to an electric double layer with opposite polarity compared to the pure flexoelectric or order-electric case. Note that tuning $\kappa_I^{-1}$ corresponds to tuning the reservoir salt concentration and \emph{not} the average salt concentration in the hedgehog defect. For no ion partitioning [Fig. \ref{fig:chargenumber}(a)], we see that higher values of flexoelectric coupling $G$ lead to higher $N$ and that the highest $N$ occurs for small $\kappa_I^{-1}$ (high values of $\rho_S^I$). When there is ion partitioning with opposite polarity, we see in Fig. \ref{fig:chargenumber}(b) that increasing $G$ initially reduces the amount of charge in the isotropic core, which is then followed by an increase in $N$ when the core changes from negatively charged to positively charged.

Finally, we note that despite the isotropic core being electrically charged, the defect as a whole, together with the nematic region, is globally charge neutral. In this sense, Fig. \ref{fig:EDLhedgehog}(c) shows that the highest local ion densities are concentrated close to the isotropic core, but for sufficiently large $r$, the screening cloud becomes oppositely charged, with a small positive (only ion partitioning) or negative (only flexoelectricity) net charge density.  In other words, there is a strong localization of charge in the isotropic core, whereas the neutralizing charge is spread out over a larger volume sufficiently far from this core in order to maintain global charge neutrality. As a practical illustration, this ``sign switch" of the net ionic charge occurs in Fig. \ref{fig:EDLhedgehog}(c) for $r\approx215\ \mathrm{nm}$ (red lines), $r\approx 60\ \mathrm{nm}$ (green lines), and  $r\approx190\ \mathrm{nm}$ (blue lines).

\section{Wedge disclinations}
\label{sec:wedge}
\begin{figure*}[t]
\centering
\includegraphics[width=\textwidth]{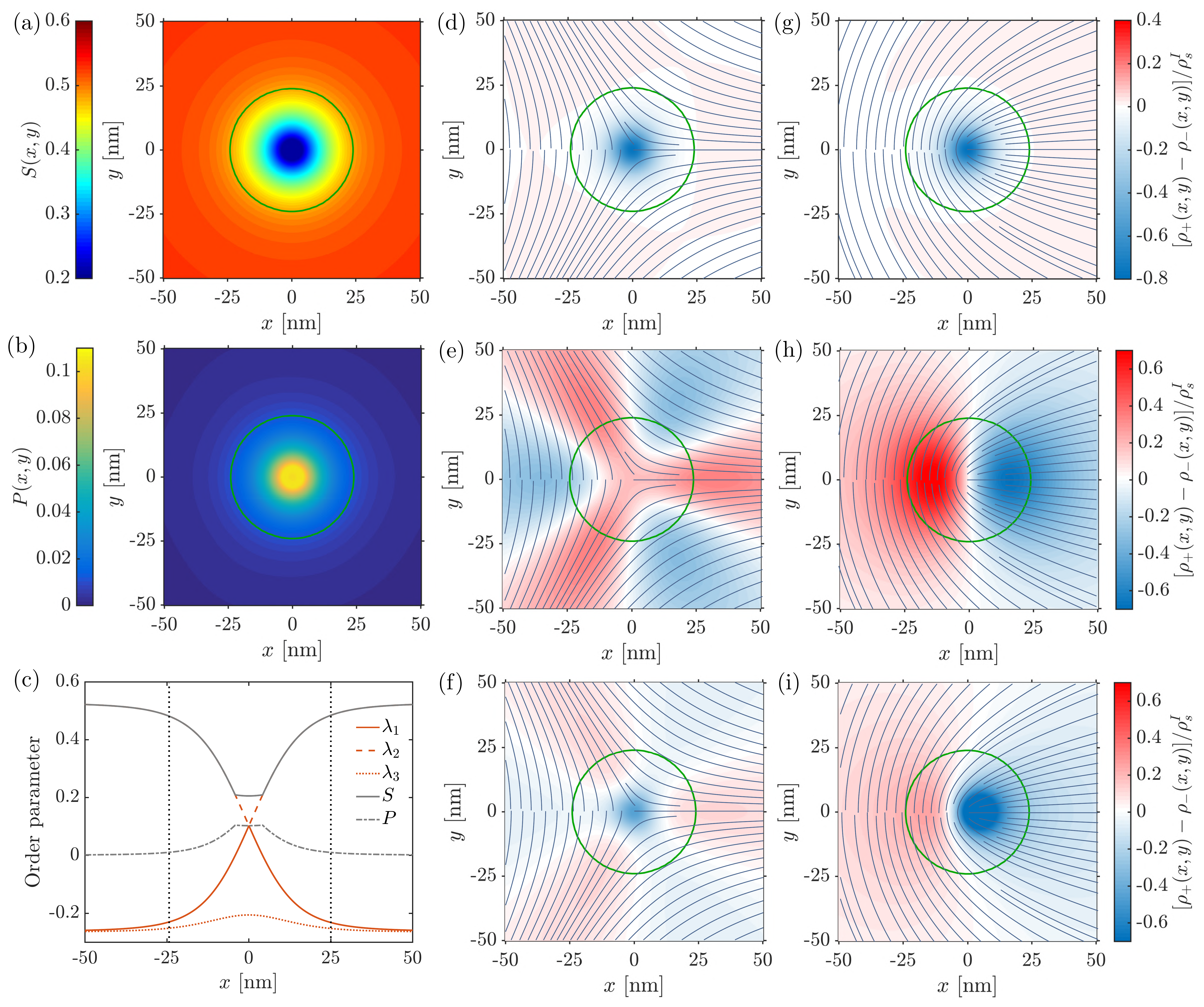}
\caption{Nematic structure and ion-charge distributions around $\pm 1/2$ wedge disclinations. (a) Scalar order parameter $S$, with the isosurface $S=0.48$ (in green) indicating the effective defect core. (b) Biaxial order parameter $P$. (c) Eigenvalues of tensor order parameter ${\bf Q}$ and scalar order parameters along $x$ for $y=0$. The dotted line is the isosurface $S=0.48$. (d)--(f) Net ion-charge distributions $[\rho_+(x,y)-\rho_-(x,y)]/\rho_s^I$ for a  $-1/2$ defect, for the cases with (d) only ion partitioning ($g_+=3$, $g_-=8$, $G=0$), (e) only flexoelectricity ($g_\pm=0$, $q_eG=10$ pC m$^{-1}$), and (f) both effects ($g_+=3$, $g_-=8$, $q_eG=10$ pC m$^{-1}$). (g)-(i) Same plots for a $1/2$ defect for (g) only ion partitioning ($g_+=3$, $g_-=8$, $G=0$), (h) only flexoelectricity ($g_\pm=0$, $q_eG=10$ pC m$^{-1}$), and (i) both effects ($g_+=3$, $g_-=8$, $q_eG=10$ pC m$^{-1}$). In all plots, we use $\kappa_I^{-1}=10\ \mathrm{nm}$, which results in an isotropic bulk ion density $\rho_s^I=1.1\times10^{-4}$ M and the streamlines in (d)-(i) indicate the nematic-director pattern.}
\label{fig:EDLwedge}
\end{figure*}

Nematic topological defect lines are another type of elementary object that can be formed in nematic electrolytes as a result of surface- or external-field-imposed frustration of the orientational order, and --as we show-- they can perform as effective diffuse capacitors. We use the full ${\bf Q}$-tensor theory in our calculations from Sec. \ref{sec:freeenergy} and exploit the translational invariance along the $z$ direction, and we impose a uniaxial far-field condition for ${\bf Q}(r\rightarrow\infty,\varphi,z)$ (in cylindrical coordinates) with bulk order parameter $S_b$ and a director field given by Oseen's solution \cite{Oseen:1933, Frank:1958, Chan:1986} 
${\bf n}({\bf r})=[\cos(k\varphi),\sin(k\varphi),0]$,
where $k$ is the winding number of the defect.


\begin{figure*}[t]
\centering
\includegraphics[width=\textwidth]{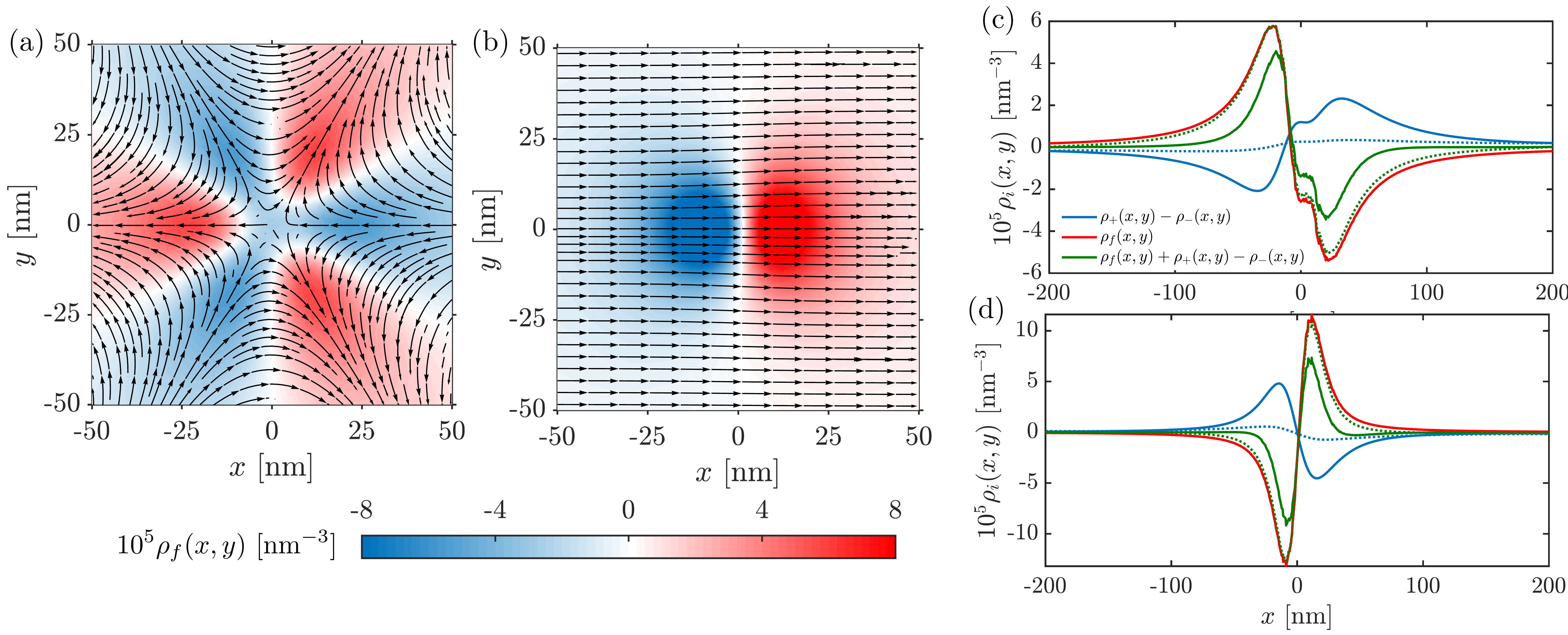}
\caption{Comparison of flexoelectric bound charge with the ion free charge density in $\pm 1/2$ wedge disclinations. (a, b) Flexoeelectric polarization charge density $\rho_f(x,y)$ as a color map, with arrows indicating the direction of the flexoelectric polarization $q_e{\bf P}_f(x,y)$ for $g_\pm=0$, $q_eG=10\ \mathrm{pC}\ \mathrm{m}^{-1}$, and $\kappa_I^{-1}=10\ \mathrm{nm}$ for (a) $k=-1/2$ and (b) $k=1/2$. (c, d) Charge profiles along the $x$ axis for the same parameter values, where we also compare with the net ion-charge density $\rho_+(x,y)-\rho_-(x,y)$ for (c) $k=-1/2$ and (d) $k=1/2$ defects. The solid lines are for $\kappa_I^{-1}=10\ \mathrm{nm}$, while the dotted lines are for $\kappa_I^{-1}=50\ \mathrm{nm}$.}
\label{fig:flexopol}
\end{figure*}


In Fig. \ref{fig:EDLwedge}, we show the results of solving the EL equations of such a system under the assumption of global charge neutrality. The nematic order parameter profiles are not influenced by the ions in this parameter regime. In Fig. \ref{fig:EDLwedge}(a), we plot the uniaxial order parameter (nematic degree of order) $S$, whereas in Fig. \ref{fig:EDLwedge}(b), we plot the biaxial order parameter $P$. The green curve in all plots indicates the (cylindrical) isosurface $S=0.48$, which effectively corresponds to the isotropic core of the defect. In Fig. \ref{fig:EDLwedge}(c), we plot the eigenvalues of ${\bf Q}$ along the cross section $y=0$, along with the values of $S$ and $P$. The dotted lines indicate the core region. For these wedge disclinations, we also find that the ${\bf Q}$ tensor is not influenced by the presence of ions or flexoelectricity, as was also the case for the flat IN interface and the radial hedgehog defect.

\begin{figure*}[t]
\centering
\includegraphics[width=\textwidth]{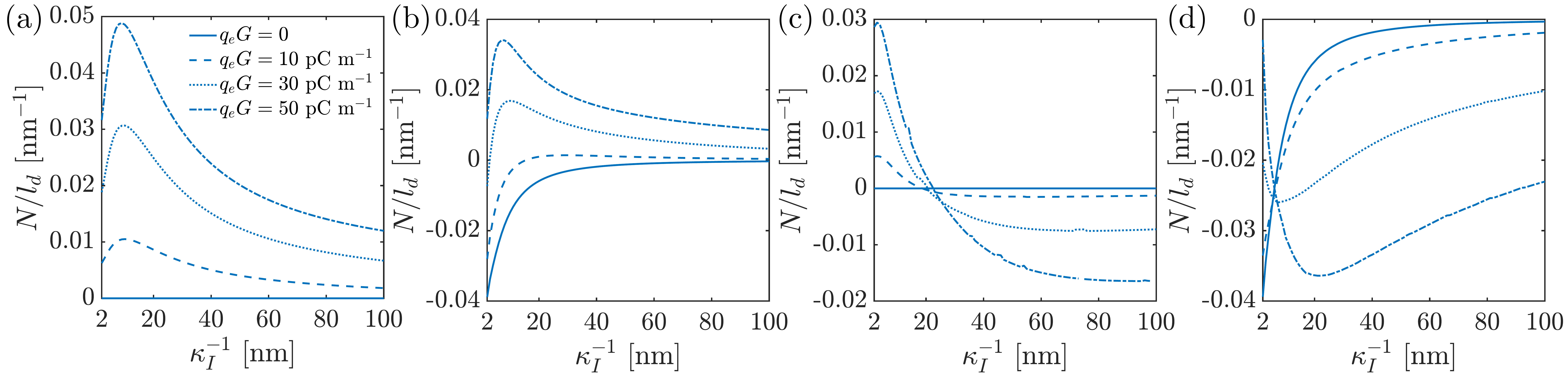}
\caption{Total charge number $N$ per unit contour length $l_d$ of half-integer wedge disclinations as a function of isotropic Debye screening length $\kappa_I^{-1}$. (a,b) For $k=-1/2$ and (c,d) $k=1/2$ defects, with (a,c) only flexoelectricity ($g_\pm=0$) with varying values of the flexoelectric coefficient $G$ and (b,d) with ion partitioning $g_+=3$ and $g_-=8$ for various values of $G$.}
\label{fig:chargenumwedge}
\end{figure*}

In Figs. \ref{fig:EDLwedge}(d)-(f), we plot the net ionic charge density $[\rho_+(x,y)-\rho_-(x,y)]/\rho_s^I$ for a $k=-1/2$ defect. The streamlines show the director profile in order to aid visualization of the defect. In Fig. \ref{fig:EDLwedge}(d), we show a representative example for $g_\pm\neq0$ but $G=0$. We see that the core is negatively charged since $g_->g_+$, although the core, in this case, is not isotropic but biaxial. The external potential governing ion partitioning, in this case, is $\beta V_\alpha({\bf r})=g_\alpha[3S({\bf r})^2/2+2P({\bf r})^2]$, which suggests that the ions couple the same way to $S({\bf r})$ as to $P({\bf r})$. However, the combination in brackets is smaller in the core, so there is still preferential ion partitioning within the (biaxial) core. The ion cloud within the core is cylindrically symmetric, but outside the core, where the double layer is positively charged, we observe that the double layer takes over the symmetry of the director profile, albeit with a much smaller (positive) charge density. 

In  Fig. \ref{fig:EDLwedge}(e), we show how the ions couple to flexoelectricity and order electricity, and we see a more complicated charge pattern. A core region with positive charge is formed with three lobes of negative charge. Therefore, we call this an electric quadlayer instead of an electric double layer. In  Fig. \ref{fig:EDLwedge}(f), we see the combined effects from { Figs. \ref{fig:EDLwedge}(d) and \ref{fig:EDLwedge}(e).} Because of the choice $g_+<g_-$, the effect is that the center of the core in Fig. \ref{fig:EDLwedge}(e) becomes negatively charged. If we would have taken $g_+>g_-$, the positive charge in the center would have been enhanced compared to the pure flexoelectric case. In Figs. \ref{fig:EDLwedge}(g)-(i), we show the same calculations but for $k=1/2$. The physics is still the same as for $k=-1/2$, but the symmetry is different. For example, when there is only flexoelectricity and order electricity, an electric double layer is formed around the core instead of an electric quadlayer [compare Fig. \ref{fig:EDLwedge}(e) with Fig. \ref{fig:EDLwedge}(h)].

\begin{figure*}[t]
\centering
\includegraphics[width=\textwidth]{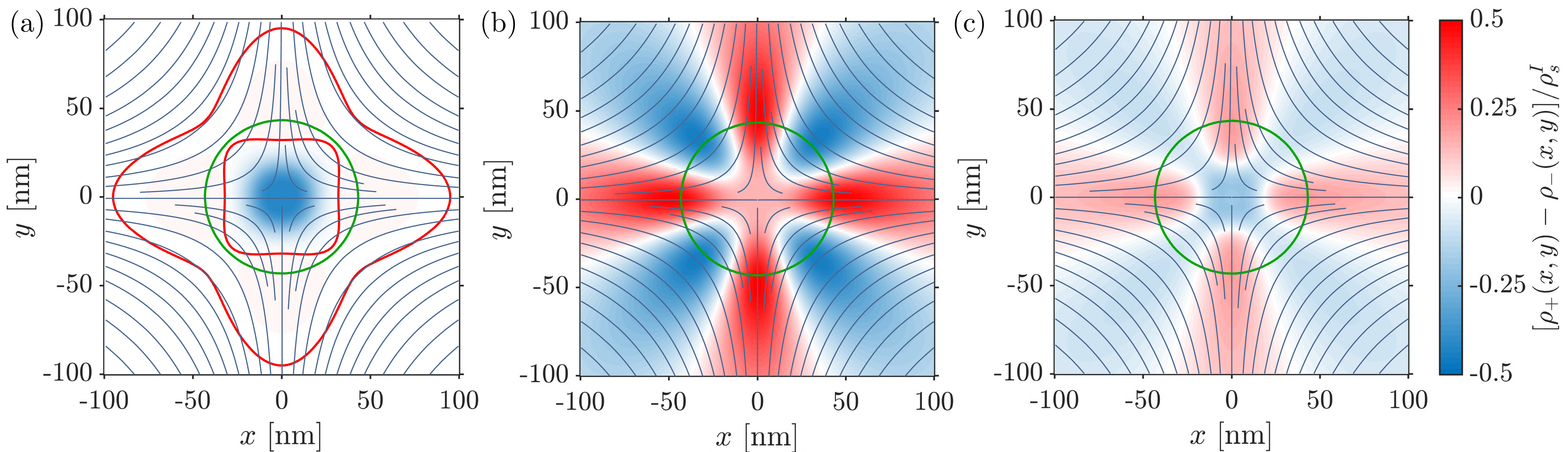}
\caption{Net ion-charge distributions $[\rho_+(x,y)-\rho_-(x,y)]/\rho_s^I$ around a $k=-1$ wedge disclination. (a) Only ion partioning, $g_+=3$, $g_-=8$, $G=0$. The dark red line is the isosurface $[\rho_+(x,y)-\rho_-(x,y)]/\rho_s^I=0.01$, indicating the low positive-charge density surrounding the negatively charged core. (b) Only flexoelectricity ($g_\pm=0$, $q_eG=10$ pC m$^{-1}$), and (c) both effects ($g_+=3$, $g_-=8$, $q_eG=10$ pC m$^{-1}$). In all plots, we use $\kappa_I^{-1}=10\ \mathrm{nm}$, which results in an isotropic bulk ion density $\rho_s^I=1.1\times10^{-4}$ M and the streamlines indicate the nematic director pattern surrounding the defect.}
\label{fig:minusone}
\end{figure*}

 The structure of the electric double layer (quadlayer) for $k=-1/2$ [Fig. \ref{fig:EDLwedge}(e)[ and the double layer for $k=+1/2$ [Fig. \ref{fig:EDLwedge}(h)] can be understood from the flexoelectric bound charge density $\rho_f({\bf r})=-\nabla\cdot{\bf P}_f=-G\partial_i\partial_jQ_{ij}({\bf r})$. Charge separation is energetically costly, and since the system wants to be locally charge neutral as much as possible, the ion profile will follow the profile of $\rho_f({\bf r})$: Where $\rho_f({\bf r})>0$, we see that $\rho_+({\bf r})-\rho_-({\bf r})<0$ and vice versa. To illustrate this case, we plot $\rho_f({\bf r})$ in Fig. \ref{fig:flexopol}(a) for the case of Fig. \ref{fig:EDLwedge}(e) ($k=-1/2$) and in Fig. \ref{fig:flexopol}(b) for the case of Fig. \ref{fig:EDLwedge}(h) ($k=+1/2$). The flexoelectric bound charge patterns are the same as the ionic patterns, but with a different sign. The streamlines in both plots show the direction of the  flexoelectric polarization $q_e{\bf P}_f({\bf r})$. 

In Figs. \ref{fig:flexopol}(c)-(d), we show the $\rho_+({\bf r})-\rho_-({\bf r})$ and $\rho_f({\bf r})$ along the $x$ axis for $k=-1/2$ in Fig. \ref{fig:flexopol}(c) and for $k=1/2$ in Fig. \ref{fig:flexopol}(d) for two different screening lengths, $\kappa^{-1}_I=10\ \mathrm{nm}$ and $\kappa^{-1}_I=50\ \mathrm{nm}$. Although the precise structure of the profiles differs for both defects, we see that $\rho_f(x,0)$ is independent of $\kappa_I^{-1}$ as we have observed earlier. For physical values of $\kappa_I^{-1}$, in our calculations, we always see that $\rho_+({\bf r})-\rho_-({\bf r})<\rho_f({\bf r})$ and that, for higher Debye screening length, the profile of $\rho_+({\bf r})-\rho_-({\bf r})$ is more spread out and smaller in amplitude. Indeed, when there are not enough ions available to locally neutralize the  flexoelectric bound charge density, ions prefer to maximize their entropy by spreading more homogeneously throughout the defect.
In other words, lower screening length means better screening, and the ions experience only the local field of the flexoelectric bound charge; thus, they adapt better to this profile, while for high screening length, ions also experience the effects of the bound charges that are farther away. Hence, a lower value of $\kappa_I^{-1}$ means stronger localization of ionic free charges around the flexoelectric bound charges. Although the amplitude might differ, the symmetry is always the same; only the charges are more spread out.

The total charge per unit contour length of the wedge disclination is shown in Fig. \ref{fig:chargenumwedge} as a function of $\kappa_I^{-1}$ for the two types of defects that we consider.  Fig. \ref{fig:EDLwedge}(e) shows that a $-1/2$ defect with only flexoelectric coupling is predominantly positively charged for $G>0$. Increasing $\kappa_I^{-1}$ means reducing the amount of ions but also less localization to the flexoelectric bound charge. If there is strong localization, there is a local minimum in the blue lobe, which reduces if $\kappa_I^{-1}$ is increased. Hence, initially, $N$ increases as well. Then, we see a decline in $N$ because $\rho_s^I$ decreases. Ion partitioning, on the other hand, always makes the core negatively charged [see Fig. \ref{fig:EDLwedge}(d)], and increasing $\kappa^{-1}_I$ will not change the shape of the profile, only how it is extended in space and its amplitude. Hence, increasing $\kappa_I^{-1}$ means that the total negative charge declines [see Fig.  \ref{fig:chargenumwedge}(b), solid line]. When flexoelectricity is turned on, these two effects are superimposed, and we get the other curves in panel (b). A similar behaviour is observed for $k=1/2$ defects [Figs. \ref{fig:chargenumwedge}(c)-(d)]. At low $\kappa_I^{-1}$, there is a strong localization of charge around the flexoelectric bound charge density; however, increasing $\kappa^{-1}_I$ results in ions getting more delocalized from the flexoelectric bound charge background, and we observe that the defect core becomes net negative charged. Ultimately,  the net negative charge in the defect should also decrease because increasing $\kappa_I^{-1}$ is equivalent to reducing $\rho_s^I$.

 \begin{figure*}[t]
\centering
\includegraphics[width=\textwidth]{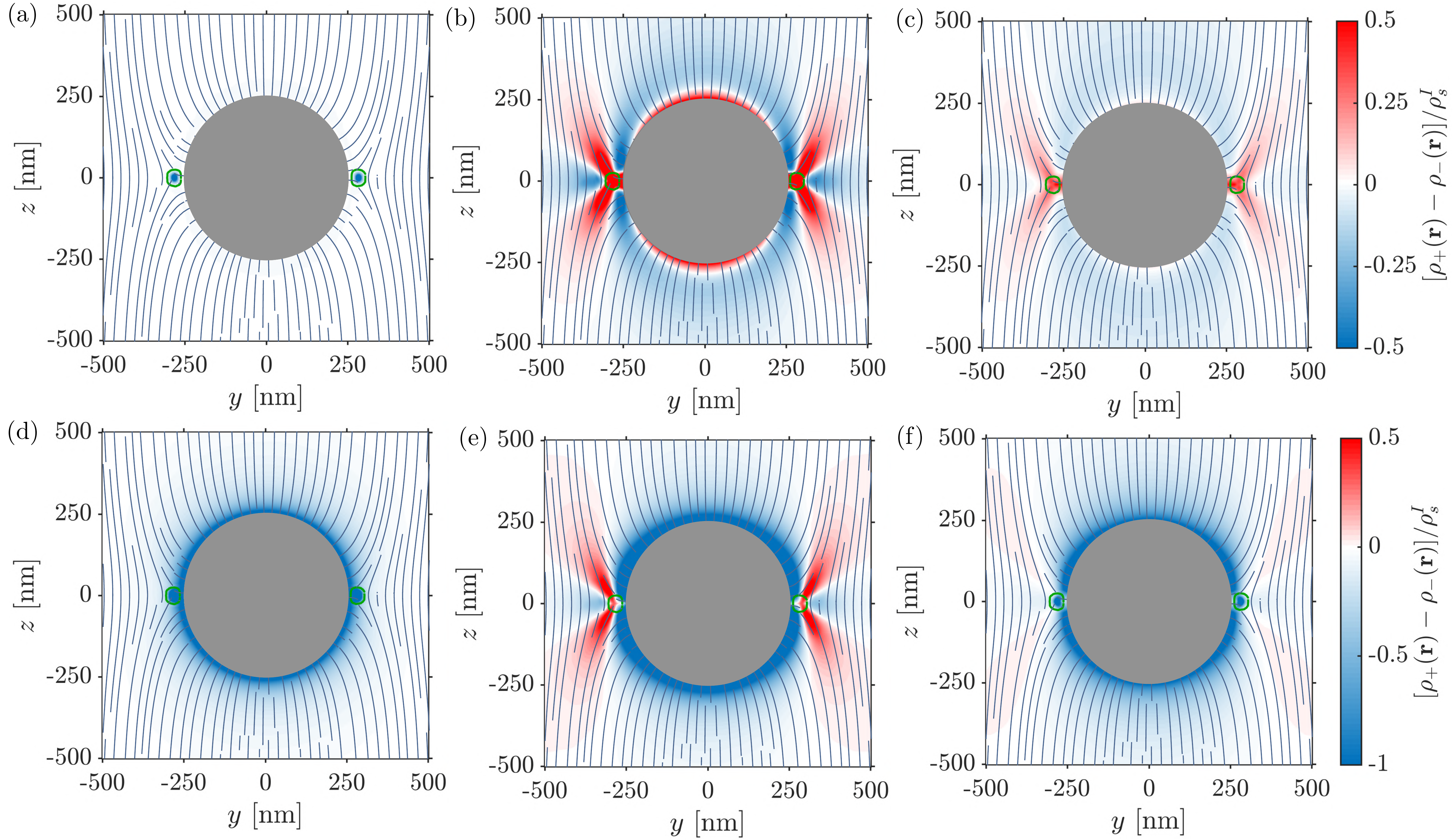}
\caption{Net ion-charge distributions $[\rho_+({\bf r})-\rho_-({\bf r})]/\rho_s^I$ around a Saturn-ring defect formed around a colloidal sphere of radius $a=250$ nm with strong homeotropic boundary conditions. The streamlines indicate the nematic-director pattern. In panels (a)--(c), the particle is uncharged, while in panels (d)--(f), the particle has a constant-surface charge density $q_e\sigma=0.001q_e \ \mathrm{nm}^{-2}$. We distinguish cases with (a,d) only ion partitioning ($g_+=3$, $g_-=8$, $G=0$). (b,e) only flexoelectricity ($g_\pm=0$, $q_eG=10$ pC m$^{-1}$), and (c,f) both effects ($g_+=3$, $g_-=8$, $q_eG=10$ pC m$^{-1}$). In all plots, we use $\kappa_I^{-1}=25\ \mathrm{nm}$, which results in an isotropic bulk ion density $\rho_s^I=1.8\times10^{-5}$ M.}\label{fig:saturn}
\end{figure*}

Finally, we show that the charge patterns that can be realized are indeed very rich and dependent on the topology of the defect. In Fig. \ref{fig:minusone}, we show the ion distributions around a $k=-1$ defect line. The effective defect core is larger for such a defect than the half-integer defects discussed above, as is indicated by the green contour lines in Fig. \ref{fig:minusone} compared to the one shown in Fig. \ref{fig:EDLwedge}. In Fig. \ref{fig:minusone}(a), we consider only the effects of ion partitioning, observing a buildup of a negative spherical-charge distribution in the  core of the defect. The neutralizing positive charge, however, is much lower, and we use red contour lines to visualize them, reflecting the different symmetry of the defect compared to the $\pm 1/2$ cases. The flexoelectric case shows a more intricate charge pattern [Fig. \ref{fig:minusone}(b)], reflecting the director profile symmetry. Compared to the $\pm 1/2$ defects where the flexoelectric case results in a double layer and a quadlayer around the defect core, the $-1$ defect features an electric pentalayer, where a positively charged region is surrounded by four lobes of negative charge. When ion partitioning is combined with flexoelectricity with $g_+<g_-$, there is a possibility of having a negatively charged region with four lobes of positive charge, as is seen in Fig. \ref{fig:minusone}(c).

\section{Liquid-crystal colloids}
\label{sec:colloid}
\begin{figure*}[t]
\centering
\includegraphics[width=\textwidth]{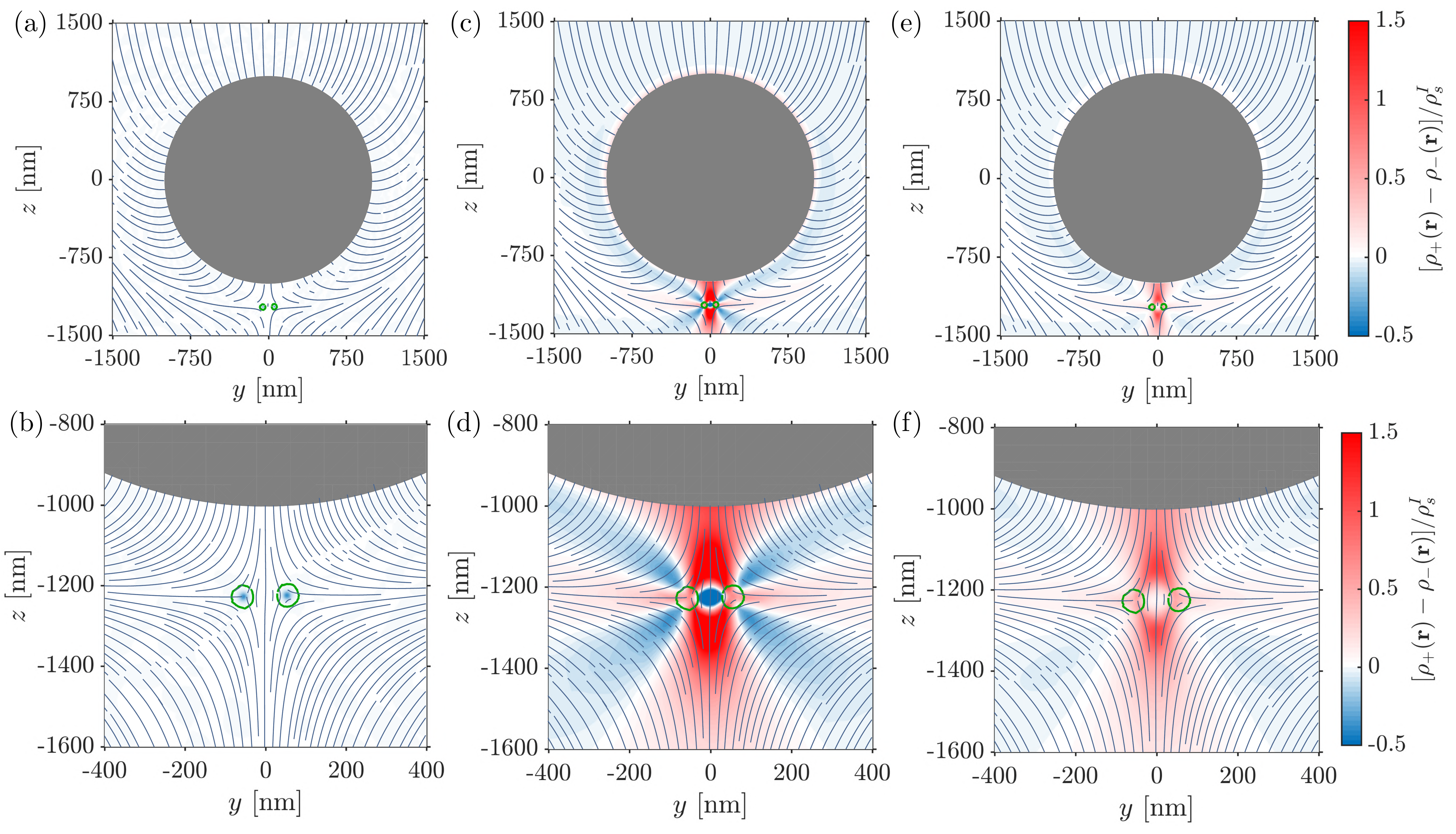}
\caption{Net-ion charge distributions $[\rho_+({\bf r})-\rho_-({\bf r})]/\rho_s^I$ around a point defect formed next to an uncharged colloidal sphere of radius $a=1$ $\mu$m with strong homeotropic boundary conditions. The streamlines indicate the nematic-director pattern. The first row shows the ion distribution around the particle; the second row shows a zoomed-in version around the defect. We distinguish cases with (a,b) only ion partioning ($g_+=3$, $g_-=8$, $G=0$), (c,d) only flexoelectricity ($g_\pm=0$, $q_eG=10$ pC m$^{-1}$), and (e,f) both effects ($g_+=3$, $g_-=8$, $q_eG=10$ pC m$^{-1}$). In all plots, we use $\kappa_I^{-1}=25\ \mathrm{nm}$, which results in an isotropic bulk ion density $\rho_s^I=1.8\times10^{-5}$ M.}
\label{fig:point}
\end{figure*}
Topological defects emerge inherently in nematic liquid-crystal colloids caused by the specific anchoring conditions at the particle surface. A central example is a spherical particle of radius $a$ with homeotropic (perpendicular) boundary conditions, while the far field has a uniform director field along, say, the $z$ direction. Depending on the anchoring strength, particle size, and elastic constant, either a point defect or a Saturn-ring defect is formed, which we show directly affects how ions are distributed around the particle. In the calculations, we assume a uniform far field and homeotropic anchoring at the particle surfaces. We assume that the particle is impenetrable for ions and has dielectric constant $\epsilon_p=2$, and constant-surface charge density $q_e\sigma$. In order to numerically stabilize the defects, we use a relaxation-type solving procedure with initial conditions based on the multipole expansion of the director field.


\begin{figure}[t]
\centering
\includegraphics[width=0.5\textwidth]{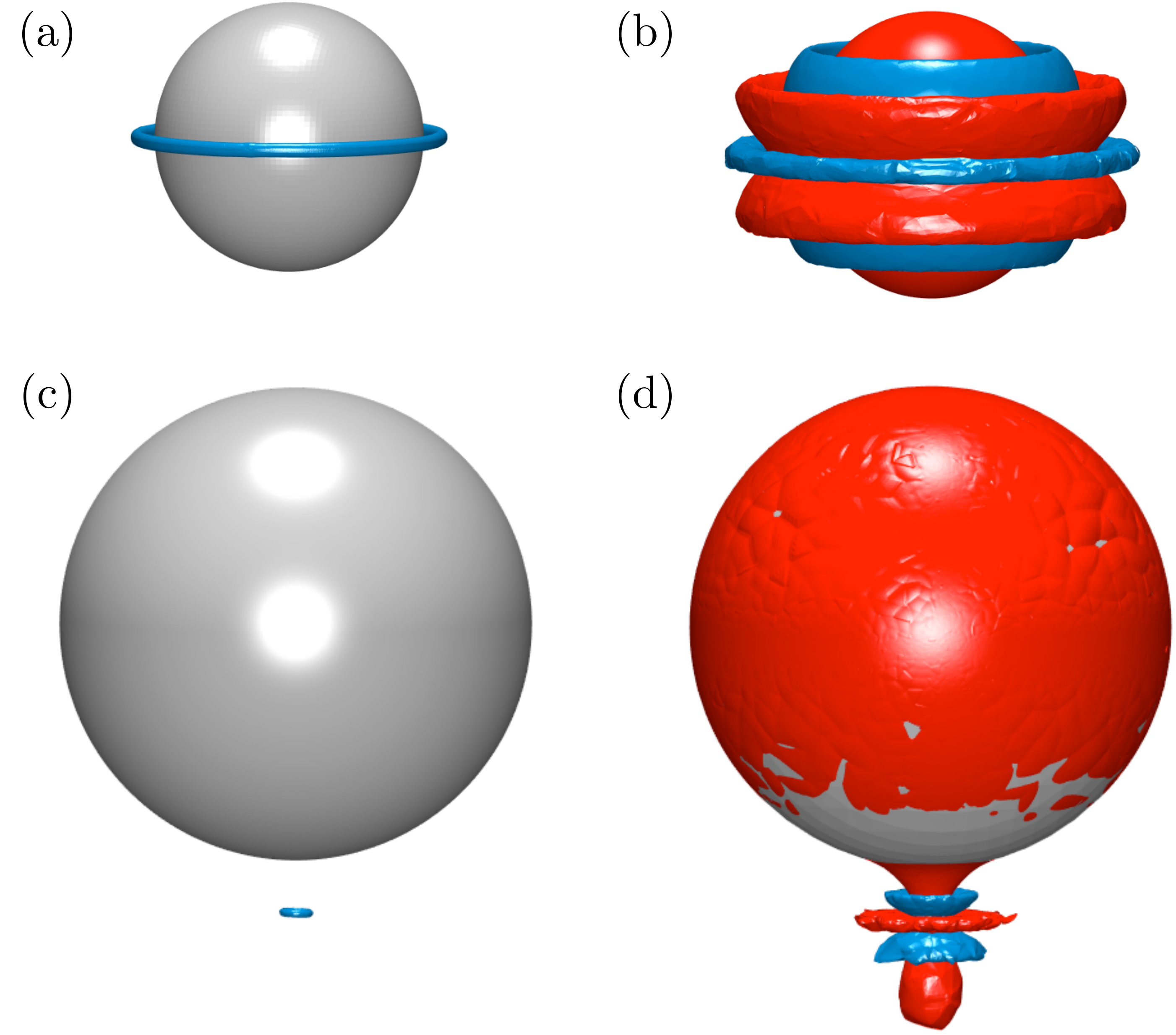}
\caption{Some selected isosurfaces of the net charge density $[\rho_+({\bf r})-\rho_-({\bf r})]/\rho_s^I$ in the screening cloud around an uncharged colloidal sphere. In panels (a) and (b), we set the particle size $a=250\ {\mathrm{nm}}$ such that a Saturn-ring defect is formed, while in panels (c) and (d), the particle size is $a=1\ \mu{\mathrm{m}}$, creating a point defect. (a,c) Only ion partitioning, $g_+=3$ and $g_-=8$, and (b,d)  only consider flexoelectric coupling, $q_eG=10\ \mathrm{pC}\ \mathrm{m}^{-1}$. The isosurface values for $[\rho_+({\bf r})-\rho_-({\bf r})]/\rho_s^I$ are (a,b)$+0.25$ (red) and $-0.25$ (blue), and (c,d) -0.15 (blue) and +0.15 (red). The colloidal particle is shown in grey.} \label{fig:3ddefect}
\end{figure}

In Fig. \ref{fig:saturn}, we show the screening cloud charge density in the case of a Saturn-ring defect around a spherical particle of radius $a=250$ nm, considering the cases with only ion partitioning, and only flexoelectricity, as well as a combination of both. First, we focus on a system where the spherical particle is uncharged, $\sigma=0$. When only ion partitioning is considered, we see a buildup of negative charge in the Saturn-ring defect [see Fig. \ref{fig:saturn}(a)]. When flexoelectricity is added, the director profile induces a specific charge pattern as discussed in Sec. \ref{sec:wedge} but only when compared in a cross-sectional flat plane. Since the Saturn-ring defect is locally the same as a $+1/2$ defect, the charge pattern in Fig. \ref{fig:saturn}(b) around the Saturn ring is identical to the one of Fig. \ref{fig:EDLwedge}(e), but it is a bit distorted because of the presence of the colloidal particle. There is also a net positive-ion-charge density around the north and south poles of the particle because of the director distortions, although the particle itself is uncharged. In Fig. \ref{fig:saturn}(c), we see the effects of ion partitioning and flexoelectricity superimposed, reducing the overall charge density. Furthermore, the positive charge densities around the north and south poles are reduced.

In Figs. \ref{fig:saturn}(d)-(f) we show the effects of surface charge density on the colloidal particle for low screening lengths $\kappa_I^{-1}=25$ nm. For such low screening lengths, the electric double layer around the particle is roughly spherical, and the charge patterns around the particle are the same as in Figs. \ref{fig:saturn}(a)-(c) but with this spherical, negatively charged double layer superimposed. For higher screening lengths, the double layer is more elongated (not shown), but then the effects of the defects are washed out (see Ref. \cite{Everts:2021}).

For larger particles, the point defect is more stable than the Saturn-ring defect, as is shown by the streamlines in Fig. \ref{fig:point}. The structure is not influenced by the low salt concentrations that we consider here. With only ion partitioning, the defect core becomes negatively charged; see Fig. \ref{fig:point}(a). If we zoom in more (the defect core is much smaller than the particle size), we see in Fig. \ref{fig:point}(b) that the defect core is actually a ring, and we have a charged ring with a radius of about $50$ nm, and the director configuration can be thought of as consisting of two $-1/2$ defects that are almost fused. 

With flexoelectricity, director distortions lead to a local net charge density around the particle, but they are much smaller than the ions that gather around the defect below the particle. In contrast, in the case of the Saturn-ring defect they are of similar magnitude, compare Fig. \ref{fig:saturn}(b) with Fig. \ref{fig:point}(c). When zoomed in on the defect, we see a charge pattern that is very similar to the one in Fig. \ref{fig:minusone}(b), but with a small negatively charged core. The reason is, again, that it is actually a small ring of locally $-1/2$ disclination -- Fig. \ref{fig:EDLwedge}(e). If the defect was a true point defect, the negative charge density in the center would disappear, and the charge pattern would be more similar to the $-1$ wedge disclination of Fig. \ref{fig:minusone} when compared in a cross-sectional flat plane. Finally, when ion partitioning is added to the flexoelectric effect, the core becomes less negatively charged, and overall, the charge density decreases [see Figs. \ref{fig:point}(e) and (f)].

Although we often compare the defects in this section with the wedge disclinations of Sec. \ref{sec:wedge}, we should keep in mind that this analogy can only be made in a cross section of the defect. In reality, one should rotate the patterns in Figs. \ref{fig:saturn} and \ref{fig:point} around the $z$ axis to get the true three-dimensional charge distribution. In order to get a feel for the three-dimensional structure of the charge distribution, we show in Fig. \ref{fig:3ddefect}, for the ion-partitioning only and flexoelectricity only cases, a few selected isosurfaces of the ion-charge distributions around a colloidal sphere.

{
\section{Experimental relevance}
We envisage three lines of possible experiments where the presented results could be realised or have a prominent role: (i) direct measurements of charge patterns, (ii) interactions and self-assembly of colloidal particles, and (iii) electrochemistry of nematic electrolytes. The vision of such work would be to create an ionically charged soft matter platform, where objects including topological defects and colloidal particles could perform as distinct microelectronic elements, capable of advanced charge manipulation, storage, and overall charge control. A clear experimental challenge in realising such a platform  is to be able to quantify and control the multiple relevant material mechanisms, including not only the elastic, dielectric, and flexoelectric properties of the nematic solvent  --which  are usually known to a fair degree-- but also the ion solvability,  the isotropic (reference) Debye length (both in the nematic and the isotropic phase), and 
the shift in the isotropic-nematic transition temperature caused by the ions (to determine $g_\pm$)
which are much less (or even poorly) known today in nematic systems.  

The predicted charge patterns could possibly be determined in direct measurements by extending the x-ray reflectivity \cite{Luo:2016} or surface plasmon resonance \cite{Chen:2017, Luo:2018} methods applied to isotropic electrolytes to nematic systems. In particular, one could possibly measure double layers by determining, for example, the Donnan potential generated over a flat isotropic-nematic interface (see Sec. \ref{sec:flat}), which already gives full insight into ionic charging of nematics but within a rather simpler planar geometry.

Ionic charging in nematic electrolytes --likely combined with elastic interactions-- can lead to a rich landscape of anisotropic attractive-repulsive interparticle potentials, which will be different from interactions of particles with just spheroidal double layers. For example, in the absence of flexoelectricity and ion partitioning, the dielectric anisotropy alone leads to anisotropic electrostatic interactions even for spherical particles in nematic solvents \cite{Everts:2021}. This result can be traced back to the spheroidal shape of the electric double layer, and it leads to a small but measurable effect in the Brownian motion of a pair of particles. As shown in this work, at sufficiently low screening lengths, the spheroidal double layer in flexoelectric nematics  gets distorted in a very anisotropic manner, leading to charge regions of the same and opposite sign compared to the particle charge [see Figs. \ref{fig:saturn}(d)-(f)], which at the level of interparticle interactions can cause the emergence of novel anisotropic attractive and repulsive directions. Furthermore, charge-regulation effects, which depend on the specific particle-charge chemistry, can additionally affect or be used to affect particle self-assembly processes \cite{Ravnik:2020}.  Finally, note that even around uncharged colloidal spheres, complex ionic charge patterns can form (Fig. \ref{fig:3ddefect}) and are again expected to affect particle interactions.

Finally, the electrochemical response of individual or structures of charged defects can prove interesting, for example, by affecting the differential capacitance or local conductivity of the system. To the best of our knowledge, besides the role of electrode topological defects in graphene-based supercapacitors \cite{Chen:2016}, the electrochemical response of individual charged nematic topological defects is, to a large degree, unknown, with selected studies in nematic liquid crystals in the absence of defects \cite{Sprokel:1973, Murakami:1996} or defect patterns in ac-driven charged nematic cells \cite{Buka:2013, Buka:2015}. Expectedly, the response of single nematic defects or their structures could be understood by constructing equivalent circuit models, both for nonoverlapping \cite{Janssen:2019} and overlapping double layers \cite{Gupta:2020}.
}

\section{Conclusions and outlook}
\label{sec:conclusions}

This work underlies the basic physical principles behind the formation of inhomogeneous ion distributions in nematic liquid crystals. The effects of ion partitioning and a composition-dependent dielectric constant can be efficiently explored within the mean-field approximation by a Landau-Ginzburg-Poisson-Boltzmann theory \cite{Onuki:2006}. We generalised such a theory to liquid-crystalline nematic order, where the local order parameter is the tensor ${\bf Q}({\bf r})$, instead of a scalar, such as in binary fluid-fluid mixtures. Compared to fluid-fluid mixtures, liquid crystals have additional structures that make them distinct, such as the presence of topological defects \cite{Kamien:2012, Smalyukh:2020}, nematic elasticity \cite{Selinger:2018}, dielectric anisotropy \cite{Foret:2006, Tojo:2009}, order electricity, and flexoelectricity  \cite{Meyer:1969, Alexander:2007, Castles:2010, Copic:2018, Kamien:2019}. Their interplay with other fields has been investigated thoroughly, such as with fluid velocity fields \cite{Giomi:2017} and electric fields \cite{Porenta:2011}. 

Our main finding is that ion partitioning leads to a buildup of ionic charge in defect cores, whereas flexoelectricity causes ionic charge separation according to the director distortions.  Consequently, we envisage that topological defects can act as ionic charge carriers or as diffuse ionic capacitors. In general, we envisage that the electric double layer can be designed when the symmetry between flexoelectric modes is broken and when more complicated director structures are considered. This precise geometry of the double layer \cite{Everts:2021} can have profound effects on the effective interaction of colloidal particles, also when combined with the particle shape \cite{Everts:2018} or with multiple colloidal particles \cite{Copar:2014}, and can have relevance in various applications, such as controlled microcargo release \cite{Kim:2018}. Another interesting extension of this work would be to consider how ions influence, for example, the interaction between two topological defects, or how the ions in nematic fluids respond to an external electric field. More complicated geometries would be interesting --such as topological defects spanning over the whole system, between particles, in confined cavities, or within memory networks, all systems that are widely studied and developed in a nonelectrostatic context.

{ Charging of topological defects could 
be further interesting  for regimes of higher ion densities --with effects beyond our work-- where the continuum Poisson-Boltzmann-type approach breaks down. In this case, one can expect, for example, Bjerrum pair formation \cite{Valeriani:2010}, overscreening and crowding \cite{Souza:2020}, and an increase of the screening length with ion concentration \cite{Smith:2016}, which are effects known in isotropic electrolytes. In particular, we expect that ions at higher densities can, for example, disrupt the liquid-crystalline structure (both profiles of the director and degree of order), which in turn can have profound back influence on the actual charge patterns. A theoretical implementation of such effects might require an explicit modeling of the solvent, rather than viewing it as a dielectric continuum that depends on the nematic order parameter, and this is a challenging problem even for isotropic electrolytes \cite{Coupette:2018}.}

Finally, in this paper, we focused on equilibrium charge structures, but the out-of-equilibrium dynamical properties would be the logical next step. Already in isotropic systems, ions have proven to be subjected to many interesting dynamical phenomena \cite{Warren:2020}, and in the liquid-crystal context, they have been investigated in terms of anisotropic ionic conductivity and an anisotropic dielectric tensor \cite{Paladagu:2017}. { A possible idea would be to extend the theory derived in Ref. \cite{Tovkach:2017} with the effects of flexoelectricity and ion partitioning included and to look for measurable signatures of the complex charge patterns presented here.} The coupling of ion degrees of freedom and nematic degrees of freedom via flexoelectricity, ion partitioning, dielectric tensor, and anisotropic conductivity may certainly lead to additional control and manipulation in these kinds of systems and possibly to the design of nanometer-sized liquid-crystal electronic elements.


\section*{\normalsize{Acknowledgements}}
J. C. E. acknowledges financial support from the European Union's Horizon 2020 programme under the Marie Skłodowska-Curie grant agreement No. 795377 and from the Polish National Agency for Academic Exchange (NAWA) under the Ulam programme Grant No. PPN/ULM/2019/1/00257. M. R. acknowledges financial support from the Slovenian Research Agency ARRS under contracts P1-0099, J1-2462 and J1-1697. The authors acknowledge fruitful discussions with S. \v Copar and S. \v Zumer. Finally, the authors would like to thank the Isaac Newton Institute for Mathematical Sciences for support and hospitality during the programme [The Mathematical Design of New Materials] when work on this paper was undertaken. This work was supported by EPSRC grant number EP/R014604/1 and COST action EUTOPIA (CA17139).

\bibliography{literature1} 

\end{document}